# Neutron Star Astronomy at ESO: the VLT decade

**Roberto P. Mignani**

*Mullard Space Science Laboratory, University College London, United Kingdom*

**Abstract**

Forty years have gone by since the optical identification of the Crab pulsar (PSR B0531+21), and 25 isolated neutron stars of different types have been now identified. Observations with ESO telescopes historically played a pivotal role in the optical studies of INSs, first with the 3.6m telescope and the *New Technology Telescope* (NTT), at the La Silla Observatory, and after 1998 with the *Very Large Telescope (*VLT), at the Paranal Observatory. In this review I summarise some of the most important results obtained in ten years of VLT observations of isolated neutron stars.

## Introduction

Isolated neutron stars (INSs) were historically discovered in the radio band (Hewish et al. 1968) as sources of beamed radiation (*pulsars)* powered by the rapidly spinning (1.5 ms-6 s) hyper-magnetised compact star[1]. Interestingly enough, however, the first INS had been already observed for 25 years in the optical band. This was the Baade and Minkowski's "south preceding star" (Baade 1942; Minkowski 1942), a quite bright object (V=16.6) at the centre of the Crab Nebula in the Taurus constellation and thought to be the nebula central star. However, it was only after the discovery of a bright radio pulsar (then named NP 0532) at the centre of the Crab Nebula (Comella et al. 1969) that the Baade and Minkowski's star was indeed recognised to be an INS and the optical counterpart of the Crab pulsar. This identification was soon after certified by the discovery of optical pulsations (Cocke et al. 1969) at the radio period (33 ms).

Forty years have gone by since the optical identification of the Crab pulsar and 25 INSs of different types have been now identified in the optical, together with the classical radio pulsars. This number can be compared with the 46 INSs detected in γ-rays by the Fermi γ-ray space telescope (Abdo et al. 2009) and with the ~90 INSs detected in X-rays (Becker 2009), while those detected in radio are ~1800 (Lorimer 2009) and obviously outnumber detections at other wavelengths. Thus, despite their intrinsic faintness, (with magnitudes down to V~28) which makes them very elusive targets, optical observations proved quite successful in detecting INSs outside the radio band. As a matter of fact, the number of optically identified INSs is comparable to that of those detected at X-ray energies at the end of the ROSAT mission (1999), which implies that optical telescopes are just one step behind with respect to current front running high-energy observing facilities, with the one step forward hopefully to be taken by the new generation 42m *European Extremely Large Telescope* (E-ELT).

---

[1] According to the magnetic dipole model (Pacini 1968; Gold 1968) the measurement of the pulsar period and period derivative yield an indirect estimate of the neutron star age, assuming that it was born spinning much faster than its present period, of its rotational energy loss, and of its dipolar magnetic field.

Observations with the ESO telescopes historically played a pivotal role in the optical studies of INSs, first with the 3.6m telescope and then with the 3.5m *New Technology Telescope* (NTT), at the La Silla Observatory (see Mignani et al. 2000 for a review). Indeed, after the Vela Pulsar (Lasker et al. 1976; Wallace et al. 1977) no other INS was identified in the optical until Bignami et al. (1988) detected a possible V~25.5 counterpart to the enigmatic γ-ray source, and putative INS, Geminga with the 3.6m telescope. This result spurred the quest for optical counterparts of other INSs, with more identifications obtained in the early 1990s, also thanks to advent of the NTT and of its new generation instruments, like EMMI and SUSI, which paved the way to observations performed with the refurbished *Hubble Space Telescope* (HST), still on going (see Mignani 2007 for a review). Indeed, one can go as far as saying that neutron star optical astronomy developed only thanks to the seminal work carried out with ESO telescopes.

The ESO contribution naturally continued with the advent of the *Very Large Telescope* (VLT) in 1998. At that time, in response to a call to the Community, G. F. Bignami, P. A. Caraveo, and I proposed to ESO optical observations of INSs as one of the possible test cases for the Science Verification of the first unit (UT1) of the VLT. Our proposal was accepted, thus acknowledging the ESO pivotal role in neutron star astronomy, and the first INS in the VLT record, the Vela-like pulsar PSR B1706-44, was observed with the Test Camera on August 17 1998. Our results were promptly published (Mignani et al. 1999) on a special issue of Astronomy&Astrophysics Letters dedicated to the outcomes of the VLT UT1 Science Verification. Remarkably, ours was historically the first *submitted* publication based on a VLT observation. Thereafter, the VLT began to sign its own contribution to the optical studies of INSs, very much like its ideal predecessor, the NTT, did in the previous decade.

It is now about ten years since our publication appeared in Astronomy&Astrophysics, ideally marking the start of the VLT era in neutron star astronomy, and much happened since then. In this review I summarise some of the most important results obtained in ten years of VLT observations of INSs. More general reviews on optical observations of INSs are presented in Shearer (2008) and in Mignani (2009a,b), to which I refer for a complete reference list.

**VLT observations of rotation-powered pulsars**

Being the first class of INSs detected at optical wavelengths, rotation-powered pulsars (simply pulsars throughout this section) were the most natural targets for VLT observations. In particular, thanks to its unprecedented collecting power, the VLT has allowed to carry out, for the first time, polarimery and spectroscopy observations for pulsars fainter than V~22, which were impossible for both the 3.6m and the NTT.

Phase-averaged polarimetry observations of young (<10,000 years old) pulsars were performed at the beginning of 1999, as a part of the commissioning of the FORS1 instrument at UT1. These observations yielded a new optical identification of a pulsar, PSR B1509-58 (Wagner & Seifert 2000), based on the preliminary measurement of polarisation at the ~10% level of its candidate counterpart. Moreover, these observations allowed to measure the phase-averaged polarisation for two more pulsars, PSR B0540-69 in the LMC (Wagner & Seifert 2000), and the Vela pulsar (Wagner & Seifert 2000; Mignani et al. 2007a), the first, and admittedly also the last, measurement ever obtained for this object. For PSR B0540-69, through the comparison of the VLT polarimetry data with high-resolution optical and X-ray observations from the HST and the Chandra X-ray Observatory, Caraveo et al. (2000) were able to correlate the map of the synchrotron emission from the surrounding supernova remnant with its polarisation structure. Thus, polarisation measurements held the potential of providing deep insights into the highly magnetised relativistic environment around pulsars. Moreover, they offer a unique test bed for neutron star magnetosphere models. For instance, Mignani et al. (2007a) showed that the measured phase-averaged polarisation of the Vela pulsar (~10%), as well as that of the Crab pulsar, is much lower than predicted by different models, which might spot possible model limitations. Interestingly, Mignani et al. (2007a) also showed, for the first time, the existence an intriguing alignment, barring possible perspective effects, between the optical polarisation direction of the Vela pulsar, the axis of symmetry of the X-ray arcs and jets observed by Chandra, the pulsar proper motion, and its rotation axis (Fig.1). This alignment, also observed in the Crab pulsar (Slowikowska et al. 2009), suggests a connection between the pulsar magnetospheric activity and its dynamical interaction with the synchrotron nebula. Strangely enough, no more polarimetry observations of pulsars with the VLT were performed thereafter.

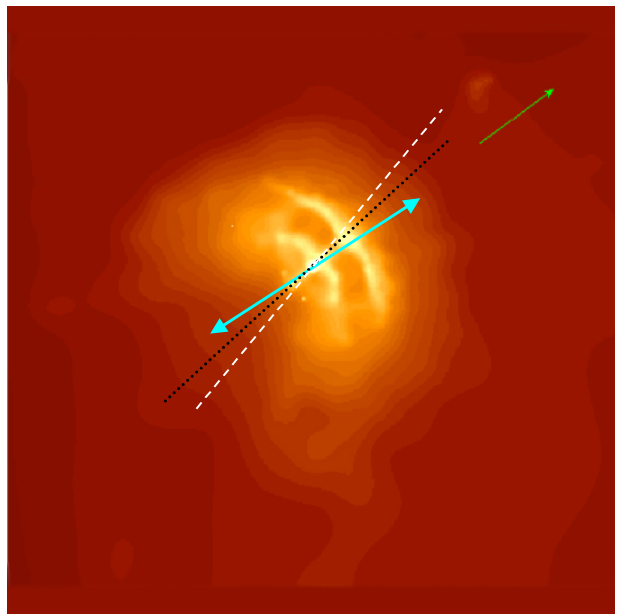

**Fig.1** *Chandra X-ray image of the Vela pulsar (centre) and its nebula (Image Credit: Chandra Press Release 26/02/2009). The white dashed and black dotted lines are the axis of symmetry of the nebula and the computed pulsar spin axis, while the green and light blue arrows are the pulsar proper motion and the polarisation direction, respectively (see Mignani et al. 2007a and references therein).*

Spectroscopy observations of the Vela pulsar, the first ever for this object, were performed with FORS1 in 2000 and 2001 (Mignani et al. 2007b). Spectroscopy is crucial to characterise the pulsar optical spectra which, for most of them, are inferred from multi-band photometry measurements, often compiled from the literature and, thus, highly inhomogeneous. The FORS1 observations showed that the optical spectrum of Vela

(4000-8000 Angstrom) is characterised by a featureless power-law continuum (Fig.2), as expected from pure synchrotron radiation (e.g. Pacini & Salvati 1983), confirming that the optical spectra of young pulsars are dominated by pure magnetospheric emission. FORS1 spectroscopy observations were also performed for another young pulsar, PSR B0540-69 (Serafimovich et al. 2004) in the LMC but, admittedly, the background from the surrounding, compact (4" diameter), supernova remnant did not allow to obtain a clean spectrum of the pulsar. The optical spectrum is more complex for middle-aged pulsars, like the ~100,000 yrs old PSR B0656+14, for which spectroscopy observations were performed, again, with FORS1 (Zharikov et al. 2007). Indeed, the optical spectrum shows the signatures of two different components: a power-law, ascribed to synchrotron radiation, like in younger pulsars, and a black body, ascribed to cooling radiation from a fraction of the neutron star surface. Similar two-component spectra are also observed in the soft X-ray emission (e.g. De Luca et al. 2005), which opens interesting possibilities to study, at the same time, spectral breaks in the synchrotron emission and to map regions at different temperatures on the neutron star surface. The study of the X-ray/optical surface thermal radiation thus offers a unique opportunity to comprehensively investigate the neutron cooling process and hence to probe the structure and composition of the neutron star interior.

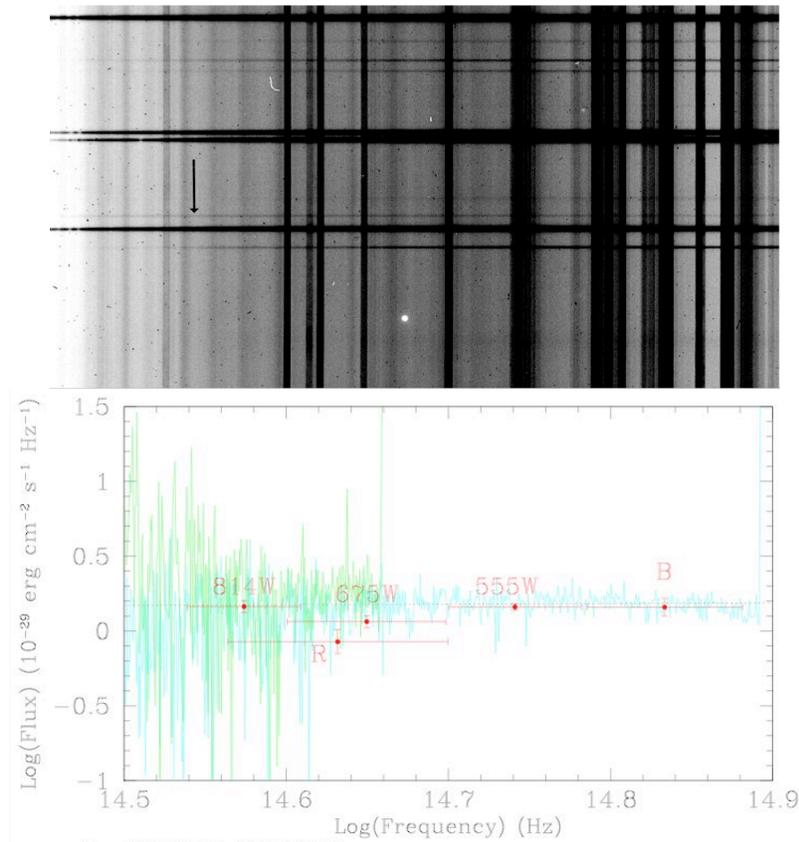

**Fig.2** *VLT/FORS1 long slit spectroscopy image (above) of the Vela pulsar (arrow) and extracted spectrum (below). Colours correspond to different grisms. Spectral fluxes from multi-band photometry are overplotted (from Mignani et al. 2007b).*

On the other hand, timing observation of pulsars, which also take advantage of the large VLT collecting area, were rarely performed, mainly due to the lack of a suited instrument. Although the HIgh Time resolution (HIT) imaging mode of FORS2 allows to perform timing observations, its time resolution (2.3-600 ms) is too low to adequately sample the light curve of fast-spinning pulsars like the Crab (33 ms), which still remains the only pulsar observed (see ESO/PR 40/99). Indeed, in most cases, timing observations of INSs at the VLT have been performed using guest instruments like, e.g. Ultra-Cam.

VLT observations with ISAAC were important to study the pulsar emission in the near IR. Multi-band (JHK) photometry of the Vela pulsar (Shibanov et al. 2002) clearly showed that the IR emission is also of synchrotron origin, like in the optical, with the spectral fluxes nicely fitting the extrapolation of the FORS1 optical spectrum (Mignani et al. 2007b). This is at variance with the younger Crab pulsar, for which ISAAC multi-band photometry (Sollerman 2003) clearly confirmed the presence of a spectral break in its optical-to-IR synchrotron emission, never observed so far in other pulsars, which might be attributed to synchrotron self absorption. The same observations allowed Sollerman (2003) to study, for the first time, the IR spectrum of the mysterious emission knot observed ~0.4" southeast of the Crab pulsar. Whether there is a connection between the pulsar and the knot is still unknown. The anti-correlation between the power-law spectra of the two objects, however, suggests that the knot is not produced by the pulsar. These results were confirmed by more recent IR observations of the Crab pulsar and its knot performed with NACO (Sandberg & Sollerman 2009).

One of the most interesting contributions of the VLT in neutron star astronomy has been in the study of the long-term evolution of the neutron star optical luminosity. According to the magnetic dipole model (Pacini 1968; Gold 1968) the pulsar luminosity is powered by the neutron star rotational energy loss. As a consequence, the luminosity of pulsars is expected to decrease due to the neutron star spin down, an effect originally predicted by Pacini & Salvati (1983) in the optical band but never convincingly measured so far. A first tentative measurement of this "secular decrease" was obtained through NTT observations by Nasuti et al. (1996) who claimed a decrement of 8±4 thousandths of magnitude per year. More recently, a new measurement has been performed by Sandberg & Sollerman (2009), also using FORS1 data, who claimed a decrement of 2.9±1.6 thousandths of magnitude per year. Although this measurement is not statistically significant, it narrows the magnitude of the effect by 60% thus imposing tighter constraints to theoretical models.

Apart from pursuing the study of optically identified pulsars, the VLT competed successfully with both the HST and with other 8m ground-based telescopes, like the Keck, the Gemini, the Subaru, and the *Large Binocular Telescope* (LBT), to obtain new identification, performing deep observations for several targets with both FORS1 and FORS2. In addition to PSR B1509-58 (Wagner & Seifert 2000), the VLT identified the likely optical counterparts of two much older pulsars: PSR B1133+16 (5 million year old) and PSR J0108-1431 (166 million years) and confirmed the proposed identification of another old (17.3 million years) pulsar, PSR B0960+08, through multi-band photometry (Zharikov et al. 2004). The optical counterpart to PSR B1133+16 was discovered thanks to FORS2 observations performed in 2003, which detected a dim source (B=28.1) at the pulsar radio position (Zharikov et al. 2008).

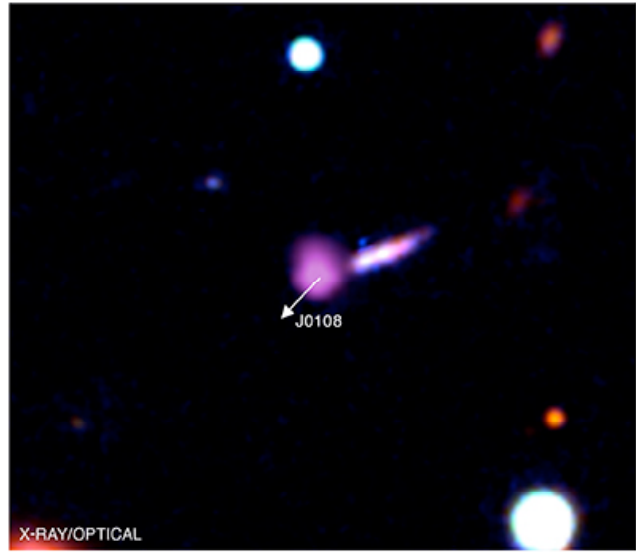

**Fig.3** *Composite VLT/FORS1 (red, blue, and white) and Chandra (purple) image of the PSR J0108-1431 field. In the X-rays, the pulsar is the source at the centre, in the optical it is the dim source north of the elliptical galaxy. The offset is due to the pulsar proper motion (arrow) between the VLT (August 2001) and Chandra (January 2007) observations (Image Credit: Chandra Press Release 26/02/2009).*

A possible counterpart (U=26.4) to PSR J0108-1431 was indeed spotted in 2001 by FORS1 observations (Mignani et al. 2003), barely detected against the halo of a nearby elliptical galaxy (see Fig. 3), but it was not recognised as such until Chandra observations discovered X-ray emission from the pulsar (Pavlov et al. 2009) and measured its proper motion with respect to the radio position. This allowed Mignani et al. (2008a) to find that the position of the candidate was consistent with the backward proper motion extrapolation of the pulsar and, thus, to certify its identification *a posteriori*. Optical/UV studies of old (>100 million year) pulsars, like PSR J0108-1431, are crucial to understand the latest stages of the neutron star thermal evolution. Indeed, these old neutron stars are expected to have cooled down to temperatures of ~10,000-100,000 K (Page 2009), making their surfaces too cold to be detectable via thermal emission in the X-ray band but still hot enough for thermal radiation to be detectable in the optical/UV bands. The detection of thermal optical/UV radiation from such old neutron stars is thus the only way to test the long-term predictions of cooling models and to investigate possible reheating mechanisms in the neutron star interior (e.g., Page 2009; Tsuruta 2009), which are more efficient in the optical/UV.

**VLT observations of other types of INSs**

Rotation-powered pulsars are only one example of the many INS types observed by the VLT. Since the 1970s, the high-energy detections of several radio pulsars, starting from the Crab and Vela pulsars, opened new perspectives in the search for INSs. In particular, some radio pulsars, for instance PSR B1509-58 and PSR B0540-69, were first detected in the X-rays and only later in the radio band. This suggested that some INSs might be more easily detectable, or possibly only manifest, at wavelengths different than radio. Peculiar types of INSs might thus be discovered, whose existence would pass unnoticed otherwise. Indeed, high-energies observations performed in the last three decades unveiled the existence of several types of INSs which are different from rotation-powered pulsars in many respects. First of all, they are typically radio-silent, while the latter are radio-loud, with the only exception of Geminga, the first INS discovered in γ-rays, and the first discovered radio-silent INS. Moreover, their multi-wavelength emission is not powered by the neutron star rotation but by other mechanisms, not yet completely understood. VLT observations of these INSs enormously contributed to the study of the nature of all of them and to determine the characteristics which make them different from rotation-powered pulsars.

**VLT observations of X-ray Dim INSs**

Observations performed mainly during the ROSAT All Sky Survey in 1990 yielded to the discovery of seven, peculiar X-ray sources with a dim (at least for the early 1990 X-ray astronomy standards) and purely thermal X-ray emission (blackbody temperatures kT~50-100 eV) and with very high X-ray-to-optical flux ratios. For this reason, these sources were immediately associated with X-ray emitting INSs located at a distance of <500 pc, as the low X-ray absorption (Hydrogen column densities $N_H \sim 10^{20}$ cm$^{-2}$) suggested. They were thus promptly nicknamed *X-ray Dim INSs*, or XDINSs (see Haberl 2007 for a recent review). The lack of magnetospheric X-ray emission, typical of young rotation-powered pulsars (Becker 2009), and of associated supernova remnants suggested that XDINSs were at least a few million year old, perhaps old enough (>500 million years) to be members of the population of once-active radio pulsars. The origin of their thermal X-ray emission was unclear, though. Depending on the actual XDINS age, it might have been originated either from a still cooling neutron star surface, or from a neutron star surface re-heated by accretion from the interstellar medium.
VLT observations were crucial to clarify the origin of the XDINS thermal X-ray emission. In particular, the identification of the optical counterpart to the XDINS RX J0720.4-3125 (B=26.7), obtained with the NTT (Motch & Haberl 1998), paved the way to the measurement of its proper motion with the VLT (Motch et al. 2003). This allowed to estimate the neutron star space velocity which, for the most likely values of the distance, turned out to be too high (> 100 km/s) to be consistent with accretion from the interstellar medium. Together with a similar conclusion inferred from the proper motion measurement of RX J1856.5-3754, obtained with the HST (Walter et al. 2001), this made XDINSs new targets to study the neutron star cooling. The measurement of X-ray pulsations (3-12 s) from all XDINSs, likely produced from hot polar caps, hinted at a possible non-uniform temperature distribution on the neutron star surface, with the colder and larger regions observable in the optical, like in the case of the middle-aged rotation powered-pulsars (see previous section). This, together with the goal of obtaining proper motion measurements for more of these objects, spurred the search for the XDINS optical counterparts, both with the HST and with the VLT. Recently, VLT observations with FORS1 and FORS2 yielded the identification of likely optical counterparts to RBS 1774 (Zane et al. 2008; Fig.4) and to RX J0420.0-5022 (Mignani et al. 2009a). Searches for XDINS have been performed also in the IR with ISAAC (Lo Curto et al. 2007;

Posselt et al. 2009) and with the Multi-conjugate Adaptive Optics Demonstrator (Mignani et al. 2008b), but with negative results so far.

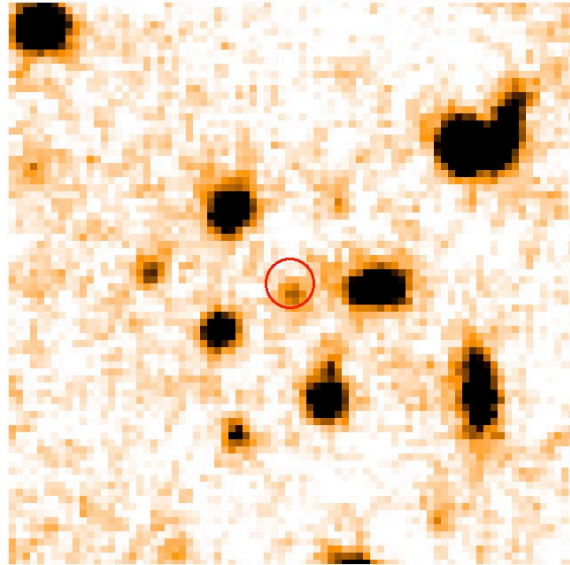

**Fig. 4.** *VLT/FORS1 B-band observation of the X-ray dim INS RBS 1774 from (Zane et al. 2008). The optical counterpart (B=27.4) is at the centre of the Chandra error circle.*

Optical spectroscopy of RX J1856.5-3754 (van Kerkwijk & Kulkarni 2001) and multi-band photometry of RX J0720.4-3125 (Motch et al. 2003), performed with FORS1, showed that their optical spectra closely follow a Rayleigh-Jeans, suggesting that their optical emission is thermal and, indeed, possibly coming from a colder and larger region on the neutron star surface. This might be true also for the other XDINSs (e.g, Mignani et al. 2009a), although the lack of optical spectral information (Mignani 2009) prevents to draw any conclusion yet. It is unlikely, however, that the XDINS optical emission, if non-thermal, is powered by the star rotation like, e.g. in young radio pulsars. Indeed, the inferred XDINS rotational energy loss is too low ($\sim 10^{30}$ erg s$^{1}$) and it would require an improbably large emission efficiency, a few orders of magnitude higher than that of radio pulsars. For RBS 1774, its relatively high optical flux is unlikely produced by the cooling neutron star surface, unless one invokes a peculiar X-ray emission geometry to explain the low pulsed fraction of the X-ray flux (Zane et al. 2008). Its optical emission may be related to the huge magnetic field ($\sim 10^{14}$ Gauss) inferred from the observation of an X-ray absorption feature (Zane et al. 2005). Confirming this interpretation, might establish a link between RBS 1774 and a family of much younger INSs: the magnetars.

### VLT observations of magnetars

*Magnetars* (about 15 known to date) represent one of the most interesting families of INSs. They have spin periods of 1.5-12 s, longer than those of most rotation-powered pulsars, and period derivatives of $10^{-13}$-$10^{-10}$ s s$^{-1}$. For a pure magneto-dipolar spin down, this corresponds to ages of ~1,000-10,000 years only and magnetic fields of ~$10^{14}$-$10^{15}$ Gauss, typically a factor of 10-100 higher than those of rotation-powered pulsars.

Historically, magnetars were identified in two families of high-energy sources (see Mereghetti 2008 for a recent review): the *Soft Gamma-ray Repeaters* (SGRs), discovered since 1979 through their recurrent soft γ-ray bursts, and the *Anomalous X-ray Pulsars* (AXPs), discovered since the 1980s through their persistent, pulsed, X-ray emission. SGRs and AXPs are thought to make a class of their own, linked by their extreme magnetic fields. According to the magnetar model (Duncan & Thompson 1992; Thompson & Duncan 1996), the torque from these extreme magnetic fields rapidly spins down the neutron star, while the field decay explain the persistent X-ray emission, much larger than it can be accounted for by the neutron star rotation, and crustal fractures induced by the field drifting explain the X/γ-ray bursts. However, other models were developed in parallel, based on accretion from low mass companion stars or from debris discs formed by the supernova explosion, which could at least explain the SGR and AXP persistent X-ray emission. Depending on the accretion regime, a disk extending to the neutron star magnetosphere might contribute to its spin down, which would mean that the inferred magnetic field values would be overestimated. Thus, SGRs and AXPs would not be *magnetars* at all but more ordinary INSs.

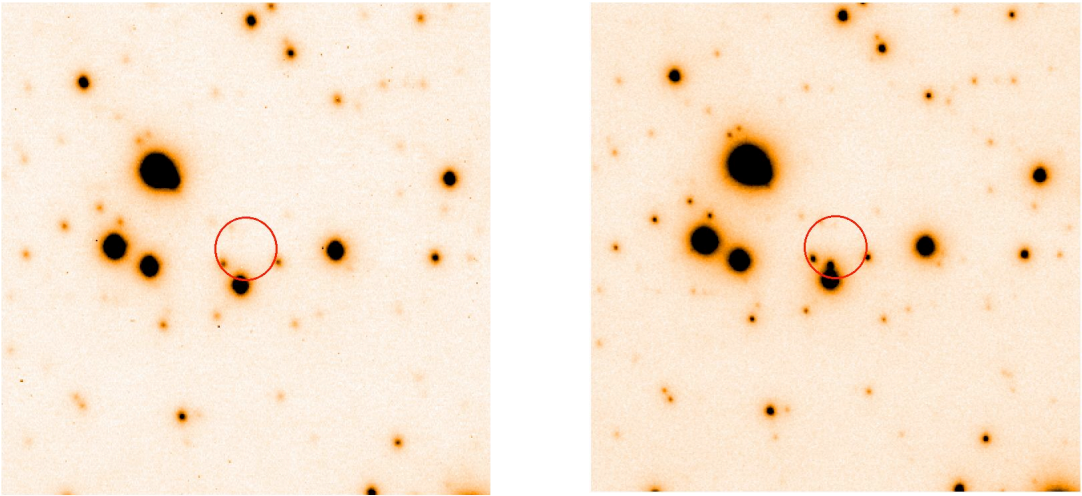

**Fig.5** *NACO Ks band images (10"x10") of the magnetar 1E 1540.0-5408 field taken in July 2007 (left; Mignani et al. 2009b) and in January 2009 (right; Israel et al. 2009). North to top, East to the left. The variable object in the error circle (0.63") is the magnetar IR counterpart.*

Optical observations were obviously crucial to test different models. Unfortunately, the field crowding towards the galactic plane, where most magnetars have been discovered, and the high interstellar extinction (up to $A_V \sim 30$) required an accurate Chandra source localisation and high-resolution IR adaptive optics observations which became feasible with the NACO instrument at the VLT. Since magnetars are variable at high energies, the best recipe to pinpoint their counterparts has relied so far on the search for sources with a correlated IR variability. This was only possible through prompt, i.e. within a few hours or days, Target of Opportunity observations in response of triggers from high-energy satellites. In this way, IR counterparts were identified for the SGR 1806-20 (Israel et al. 2005), and possibly for SGR 1900+14 (Testa et al. 2008), and for the AXPs

1E 1048.1-5937 (Israel et al. 2002), XTE J1810-197 (Israel et al. 2004), 1E 1540.0-5408 (Fig. 5; Israel et al. 2009), and possibly for 1E 1841-045 (Testa et al. 2008), amounting to most of the magnetars identified in the IR. These observation ruled out the presence of both a companion star and of a disc extending down to the magnetospheric boundary, thus supporting the magnetar model.

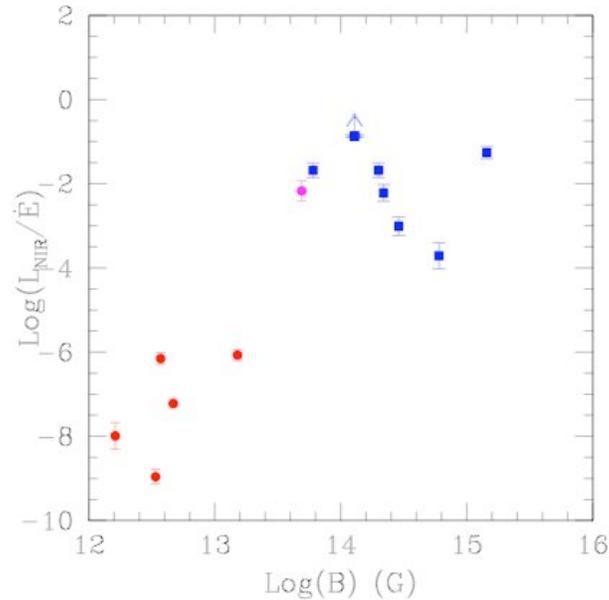

**Fig. 5** *IR luminosity/rotational energy loss ratio as a function of the magnetic field (updated from Mignani et al. 2007c) for rotation-powered pulsars (red), magnetars (blue), and rotating radio transients (purple).*

The origin of the magnetar IR emission, however, is still uncertain. Mignani et al. (2007c) showed that the ratio between the magnetar IR luminosity and the rotational energy loss is a factor of >100 larger than for radio pulsars, suggesting that it is not powered by rotation but, rather, by their larger magnetic fields (Fig. 6). Alternatively, the IR luminosity could be due to reprocessing of the X-ray radiation in a passive disc, which might have been detected in the mid-IR by Spitzer around the AXPs 4U 0142+61 (Wang et al. 2006) and 1E 2259+585 (Kaplan et al. 2009). Determining the source of the magnetar IR emission would thus represent an important test to both the magnetar and supernova explosion models. Unfortunately, the NACO phase-averaged upper limit on the polarisation of the two AXPs 1E 1048.1-5937 and XTE J1810-197 can not help to discriminate between the two scenarios (Israel et al., in preparation). At the same time, the detection of optical pulsations from the AXP 1E 1048.1-5937, obtained with UltraCam at the VLT for the first time (Dhillon et al. 2009), and the study of the light curve profile could not provide conclusive evidence in favour of a pure magnetospheric emission and against the disc reprocessing scenario. This, however, has been found to be incompatible with the uncorrelated IR-to-X-ray variability of the AXPs XTE J1810-197 (Testa et al. 2008).
IR observations might also help to clarify the link between magnetars and other INS types. In particular, the detection of transient radio emission from the AXPs XTE J1810-

197 and 1E 1540.0-5408 (Camilo et al. 2007a,b) suggests that magnetars might be related to the recently discovered class of radio-transient INSs called Rotating Radio Transients, or RRATs (see Mc Laughlin 2009 for a review). These sources (about 20 known to date) feature extremely bright radio bursts lasting only 2-30 ms which tend to repeat at interval of minutes of hours, and they have periods of 0.4-7 s, not far from those of the magnetars. Interestingly enough, NACO observations pinpointed a possible counterpart to the RRAT J1819-1458 (Rea et al. 2009), that with the highest inferred magnetic field, whose IR emission very well fits the magnetar characteristics (see Fig. 4) and might strengthen a possible link between the two INS classes.

**VLT observations of Central Compact Objects in supernova remnants**

Very much at a variance with magnetars are another type of radio-silent INSs (about ten known to date) at the centre of young (~ 2,000-40,000 year old) supernova remnants. Originally discovered by the Einstein observatory, more of these sources were discovered by ROSAT, although they obtained the lime light after the spectacular discovery of one new source at the centre of the Cas A supernova remnant with Chandra. The nature of these sources, a.k.a. central compact objects, or CCOs (see De Luca 2008 for a recent review), is puzzling. Their association with young supernova remnants suggests that they are young INSs although they feature no evidence of magnetospheric activity, either in the form of power-law X-ray spectra or of pulsar wind nebulae like, e.g those observed around the Crab and Vela pulsars. Only for three CCOs X-ray pulsations have been detected (0.1-0.4 s) with period derivatives $< 10^{-13}$ s s$^{-1}$ which imply magnetic fields $<5 \times 10^{11}$ Gauss. CCOs might thus be neutron stars born spinning close to their present period and with very low magnetic fields, which might have accreted from a debris disc.

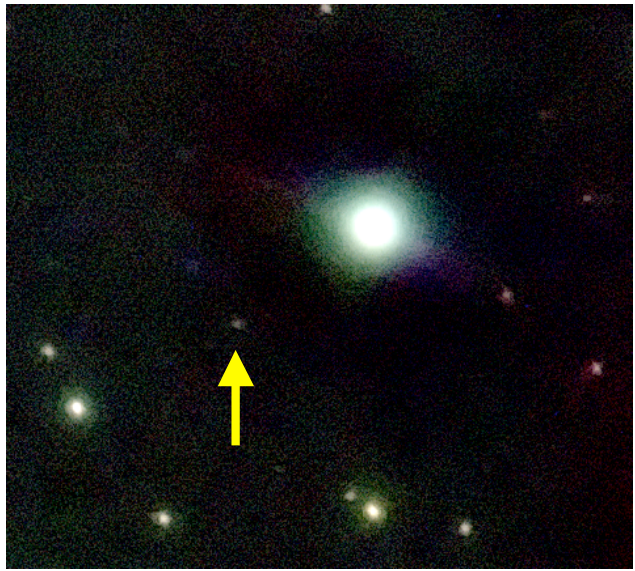

**Fig. 7**. *Composite HKs VLT/NACO image around the Central Compact Object in the Vela Jr. supernova remnant. The candidate IR counterpart (Mignani et al. 2007d) is the faint object (Ks=21.4) marked by the arrow (Image Credit: Simone Zaggia, INAF-Osservatorio Astronomico di Padova).*

Deep optical/IR observations of CCOs have been performed mainly with the VLT, with FORS1 and NACO (Mignani et al. 2008c; Mignani et al. 2009c), but no likely candidate counterpart was found. Only for the CCO in the Vela Jr. supernova remnant, a candidate IR counterpart was identified by NACO (Mignani et al. 2007d; Fig.7), whose nature, however, is yet undetermined. Interestingly, the Vela Jr. CCO also features an optical nebulosity detected both in the R band and in Hα which might be a possible bow-shock, produced by the neutron star motion in the interstellar medium. Most puzzling of all, is the source in the RCW 103 supernova remnant, which is very much different from the other CCOs, featuring a 6 hour X-ray period and spectacular long-term X-ray flux variations. It might be a magnetar, braked by a magnetic field $>10^{15}$ Gauss, or a very rare example of a neutron star in a binary system born in a supernova remnant. A search for correlated periodic or long-term X-ray and IR variability (De Luca et al. 2008), however, did not pinpoint a likely counterpart, with the upper limits being consistent with the presence of both a very low mass (>M5) star or with a debris disc. Target of Opportunity observations with NACO, following a source reactivation detected by XMM-Newton, will hopefully lead to its counterpart identification and to clarify the puzzling nature of this source.

**Conclusions**

Optical/IR observations with the VLT (see Tab. 1 for a qualitative summary) have played, and still plays, a major role in the characterisation of the multi-wavelength phenomenology of radio pulsars, the only astrophysical objects, together with AGN, which are detectable from radio to high-energy γ-rays, and in the understanding of the nature of the many peculiar INS types, like the magnetars. Moreover, VLT observations have provided important contributions to the study of the intrinsic properties of neutron stars like, e.g. the structure and composition of the neutron star interior, through the study of optical thermal radiation from the neutron star surface, and to the study the emission processes in their hyper-magnetized magnetospheres, through the study of non-thermal optical/IR radiation.

While in the years to come the VLT will still be a major leading facility in neutron star optical astronomy, its role must be seen in perspective as a pathfinder for future observations to be performed with the 42m *European Extremely Large Telescope* (E-ELT), which is expected to be operational in about ten years from now. Proceeding on the VLT beaten path, the E-ELT will be able to yield about a hundred new INS identifications, thus reducing the current gap between optical and high-energy observation (Becker 2009; Abdo et al. 2009). Many of them will likely come from the follow-up of observations with the new generation of Mega Structure facilities, like the Square Kilometer Array (SKA) in the radio band. Moreover, the E-ELT will allow astronomers to easily carry out observations (spectroscopy, and possibly, also timing and polarimetry) which, for the faintest INSs, still represent a challenge for the VLT, enabling more in-depth investigations which will eventually complete and expand the VLT seminal work.

Forty years after the identification of the Crab pulsar, the optical study of INSs is still a very active research field, where ESO marked important milestones in the last 20 years, with the NTT first (Mignani et al. 2000) and now with the VLT. The E-ELT will be able to take up this legacy, opening a third new era in neutron star astronomy at ESO.

| Name | Age | Mag | D(kpc) | $A_V$ | Instruments | Mode |
|---|---|---|---|---|---|---|
| Crab | 3.10 | 16.6$^V$ | 1.73 | 1.6 | FORS1/ISAAC/NACO | IMG,HIT-I |
| **B1509-58** | 3.19 | 25.7$^R$ | 4.18 | 5.2 | FORS1 | IPOL |
| B0540-69 | 3.22 | 22.0$^V$ | 49.4 | 0.6 | FORS1 | LSS |
| Vela | 4.05 | 23.6$^V$ | 0.23 | 0.2 | FORS1/ISAAC | IMG, IPOL, LSS |
| B0656+14 | 5.05 | 25.0$^V$ | 0.29 | 0.09 | FORS1 | LSS |
| **B1133+16** | 8.89 | 28.0$^B$ | 0.35 | 0.12 | FORS1 | IMG |
| B0950+08 | 7.24 | 27.1$^V$ | 0.26 | 0.03 | FORS1 | IMG |
| J0108-1431 | 8.3 | 26.4$^U$ | 0.2 | 0.03 | FORS1 | IMG |
| **RX J0720.4-3125** | 6.27 | 26.7$^V$ | 0.30 | 0.3 | FORS1/FORS2/ISAAC | IMG |
| **RBS 1774** | 6.57 | 27.2$^B$ | 0.34 | 0.18 | FORS1/FORS2 | IMG |
| RX J1856.5-3754 | 6.60 | 25.7$^V$ | 0.30 | 0.12 | FORS1/ISAAC/MAD | IMG, LSS |
| **RX J0420-5022** | - | 27.5$^B$ | 0.35 | 0.07 | FORS1/FORS2/MAD | IMG |
| 1E 1547.0-5408 | 3.14 | 18.2$^K$ | 9 | 17 | NACO | IMG, IPOL, LSS |
| SGR1806-20 | 3.14 | 20.1$^K$ | 15.1 | 29 | ISAAC/NACO | IMG |
| 1E 1048.1-5937 | 3.63 | 21.3$^K$ | 3.0 | 6.10 | NACO/Ultra-Cam | IMG, IPOL,HIT-I |
| **XTE J1810-197** | 3.75 | 20.8$^K$ | 4.0 | 5.1 | NACO | IMG, IPOL |

**Tab.1** *VLT detections for all INSs with optical counterparts. INSs are sorted according to the spin-down age (in logarithmic units). Different INS types are colour-coded (yellow: radio pulsars; orange: XDINSs; green: magnetars). Magnitude superscripts indicate the pass band. INS identified by the VLT are in bold. Used instruments and observing modes (IMG=imaging; IPOL=imaging polarimetry; LSS=long slit spectroscopy; HIT-I=high time resolution imaging) are reported in the last two columns.*